# Diffraction from static potential in QED

Andrew Daniels,[1] Jordan Kupec,[2] Timothy Baker,[1] Eduardo V. Flores[1]

[1]Department of Physics, Rowan University, Glassboro, NJ
[2]Department of Physics, College of the Holy Cross, Worcester, Ma

We model a wire as a cylindrical potential barrier and calculate diffraction from that potential using quantum electrodynamics. We compare our results with classical Fraunhofer diffraction. We find general agreement between the quantum and the classical results; however, we find that the classical approach overestimates the wire radius. We consider an incoming electron beam diffracting from the potential. We also consider the case of an incoming photon beam. For the photon case we only indicate the amplitudes that need to be evaluated numerically. We also study the case of two beams in phase that interfere and diffract from the wire-like potential.

## I. INTRODUCTION

Classical Optics has developed different tools to calculate diffraction for different setups [1]. On the other hand, beyond path integral approaches for some particular examples [2], one cannot find in the literature examples of diffraction calculations using the Feynman diagram approach in quantum electrodynamics (QED). The reason is that QED deals with interactions among elementary particles and an obstacle such as a wire contains a macroscopic number of elementary particles. However, if the macroscopic object could be at least partially modeled by static potential then the situation changes.

In this paper we consider diffraction, or more accurately scattering, of electrons and photons from a potential barrier using Feynman diagrams [3,4]. We model a wire as a cylindrical potential barrier. This model ignores important surface properties of an actual wire. It is not clear how to physically generate such a potential. Nevertheless, the key aspect of this potential is its geometrical similarity to the wire. Scattering of electrons from static potentials is a simple interaction at the lowest order in QED. Scattering of photons from static potentials is also possible but it is complicated by the fact that photons only interact indirectly with other photons; here we indicate the amplitudes that need to be evaluated using numerical methods.

It is possible to interpret a diffraction pattern as a probability distribution; this approach is probably the easiest way to compare the results from QED with classical results. In QED the amplitude or transition matrix may be obtained from Feynman diagrams. We multiply the absolute square of the amplitude by the number of final states and obtain the transition rate. The transition rate is proportional to a probability distribution. We compare the normalized probability distribution from QED with a normalized classical diffraction distribution using Fraunhofer approach [1].

## II. SINGLE BEAM CASE

Consider a beam of electrons travelling along the $z$-axis and a long wire centered at the origin along the $y$-axis as in Fig. 1a. In QED the lowest order contribution to electron scattering in terms of Feynman diagrams is shown in Fig. 1b.

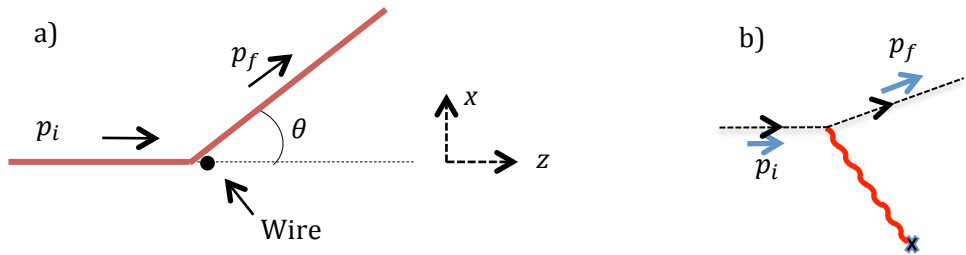

**FIG. 1. Setup for electron scattering.** a) An electron moving along the $z$-axis is deflected by a wire located at the origin along the $y$-axis. b) This Feynman diagram represents an electron scattering from an external field.

The amplitude for the process in Fig. 1b [3] for the case $f \neq i$ is represented by

$$S_{fi} = -ie \int d^4\vec{x}\, \bar{\psi}_f(\vec{x})\gamma^\mu A_\mu(\vec{x})\psi_i(\vec{x}), \quad (1)$$

where $\psi_i$ represents an incoming electron, $\psi_f$ represents the scattered electron, $A_\mu$ represents the virtual photon exchanged with the wire. In Eq. (1), the initial and final states of the electron are represented by plane waves $\psi(x) = au(p,s)e^{-ip\cdot x}$, where $a$ is a normalization constant, $p$ is the 4-momentum, $s$ is the spin, $u$ is a spinor and $x$ is the space-time coordinate. The dot product is define by $p\cdot x = \frac{E}{c}ct - \vec{x}\cdot\vec{p}$. The static field is

$$A_\mu = \begin{cases} \frac{V(\vec{x})}{c} & \mu = 0 \\ 0 & \mu \neq 0 \end{cases}. \quad (2)$$

Substituting Eq. (2) in Eq. (1) we get

$$S_{fi} = -i\frac{e}{c}a^2\bar{u}\gamma^0 u \int d^4x\, V(\vec{x}) e^{i(p_f - p_i)\cdot x}. \quad (3)$$

We model the wire as a potential barrier:

$$V(\vec{x}) = \begin{cases} H & 0 \leq s \leq R \\ 0 & s > R \end{cases}. \quad (4)$$

The potential in Eq. (4) is time independent, thus, integrating Eq. (3) over time yields the delta function $2\pi\delta(E_f - E)$. Similarly, independence of the potential on the $y$-coordinate yields the delta function $2\pi\delta(p_{fy} - p_{iy})$. Integrating the potential in Eq. (3) over the remaining coordinates gives

$$\frac{1}{\pi R^2}\int_0^R \int_0^{2\pi} s e^{-i|q|s\cos\phi'}\, d\phi'\, ds = {}_0F_1, \quad (5)$$

where ${}_0F_1\left[2, -\frac{1}{4}|q|^2 R^2\right]$ a hypergeometric regularized function of the wire radius ($R$) and the momentum transfer $q = |\vec{p}_f - \vec{p}_i|$. The amplitude in Eq. (3) is now

$$S_{fi} = -i\frac{e}{c}a^2 H\pi R^2 \bar{u}\gamma^0 u \left[2\pi\delta(E_f - E)\right]\left[2\pi\delta(p_{fy} - p_{iy})\right] {}_0F_1. \quad (6)$$

The transition probability per particle is $d\sigma = |S_{fi}|^2 \frac{d^3 p_f}{(2\pi)^3} L^3$, where $L$ is the side of normalization box. Using the identities:

$$\left[2\pi\delta(E_f - E)\right]^2 = 2\pi T\delta(E_f - E), \quad (7)$$
$$\left[2\pi\delta(p_{fy} - p_{iy})\right]^2 = 2\pi L\delta(p_{fy} - p_{iy}),$$

where $T$ is the integration time [4], we obtain the transition probability

$$d\sigma = C|\bar{u}\gamma^0 u|^2 \delta(E_f - E)\delta(p_{fy} - p_{iy}) {}_0F_1^2 d^3 p_f, \quad (8)$$

where $C$ is a constant. The delta functions require: $E_f = E$ and $p_{fy} = p_{iy}$. We set $p_{ix} = p_{iy} = 0$, $p_{iz} = p$, $p_{fz} = p\cos\theta$, and $p_{fx} = p\sin\theta$. We note that $q$ simplifies to $q^2 = 4p^2 \sin^2\left(\frac{\theta}{2}\right)$. Integrating Eq. (8) we get

$$d\sigma = C|\bar{u}\gamma^0 u|^2 {}_0F_1^2 d\theta. \quad (9)$$

We now calculate $\bar{u}\gamma_0 u$. Lets assume incoming electron beam is polarized, spin up. If the initial spin is up and the final spin is down we get

$$\bar{u}\gamma_o u = \frac{c^2 p^2 \sin\theta}{E+m c^2}. \quad (10a)$$

If the initial and final spins are up we get:

$$\bar{u}\gamma_o u = \frac{(E+m c^2)^2 + c^2 p^2 \cos\theta}{E+m c^2}. \quad (10b)$$

The probability distribution in Eq. (9) is valid at all energies. Our interest is in diffraction at energies low compared to the electron mass, $pc \ll mc^2$. In this limit Eq. (10a) is zero, Eq. (10b) is a constant and Eq. (9) becomes

$$\frac{d\sigma}{d\theta} = C \left\{ {}_0F_1 \left[2, -R^2 p^2 \sin^2\left(\frac{\theta}{2}\right)\right]\right\}^2. \quad (11)$$

In QED we are normally interested in scattering and we pay less attention to particles that go through unchanged. This is not the case in classical diffraction where we obtain the full diffraction pattern. However, using a semi-classical approach to diffraction it is possible to separate the photons that are deflected from those that go through unchanged. The final outcome is that wire scattering effects from QED can be compared with classical slit diffraction (for details see previous version of this paper). Thus, we compare Eq. (11) with Fraunhofer diffraction for the slit [1], which is proportional to

$$\text{sinc}^2(Rp \sin\theta). \quad (12)$$

We note that QED treats the wire as a 3D object while in Fraunhofer approach the wire is a 2D object [6]; this might be the reason behind an overestimation of the wire radius [7] in Fraunhofer diffraction as shown in Fig. 2.

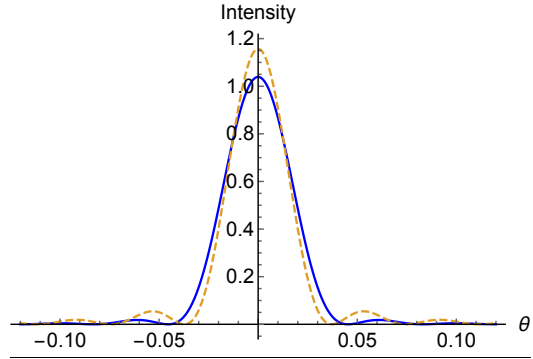

**FIG. 2. Classical and quantum probability distribution for wire diffraction.** The orange dashed curve represents classical wire diffraction. The solid blue curve represents quantum diffraction. The net area under each graph is the same. We note that first dark point for the quantum case is shifted outward from the classical case, which indicates that Fraunhofer approach overestimates the wire diameter compared to QED approach. The wire diameter is 17 $\mu$m. The electron wavelength is 633 nm.

We note that if, in the QED calculation, we increase the wire radius by a factor of 1.21 the first dark spots in Fig. 2 line up. Experimentally, we see this discrepancy with photons. In fact, direct measurement of the wire diameter gives a value of 14 $\mu$m while an indirect measurement obtained by considering the wire as a 2D strip gives 17 $\mu$m, which is 1.21 times larger than the actual value. This observation implies that, at low energies, photon diffraction results should be closer to electron diffraction in QED than to classical diffraction.

Photon diffraction from a static potential is scattering of light by light [8,9]. The lowest level contribution comes from a set of six Feynman diagrams [8,9,10] as the one in Fig. 3a.

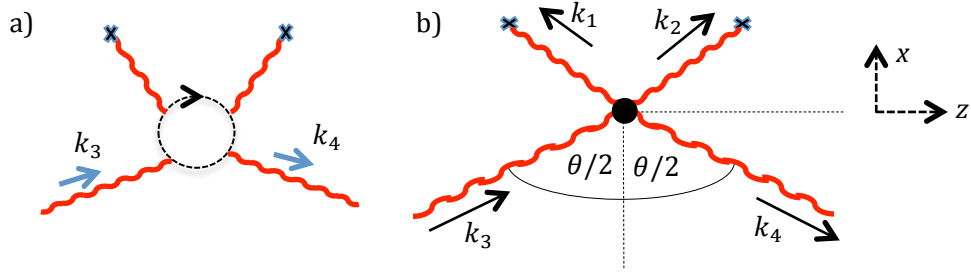

**FIG. 3. Photon scattering from static potential**. a) The photon interacts with the potential through a virtual electron loop. The **x** at the end of the photon line indicates interaction with the potential. b) Notation: the incoming photon has 4-momentum $k_3$ and the scattered photon has 4-momentum $k_4$. The scattering angle is $\theta$. The wire is placed along the $y$-axis. The 4-momenta $k_1$ and $k_2$ are the momenta exchanged with the wire.

In QED the process in Fig. 3a is closely related to Delbruck scattering first observed in 1973 [11]. In Delbruck scattering, a photon is scattered by the Coulomb potential of the nucleus of an atom. With some modifications in notation, shown in Fig. 3b, we adopt the results for Delbruck scattering in Ref. 10. This is possible because the authors of Ref. 10 approach to Delbruck scattering is general. Assuming our results follow the same overall pattern as the results for Delbruck scattering, namely, forward scattering that maintains polarization is proportional to a positive power of the cosine of scattering angle then photon diffraction at low energy turns out to be nearly identical to electron diffraction at low energy. This is an expected result due to matter-radiation symmetry for wave effects.

The amplitudes for scattering circular polarized photons from a static potential [9,10] can be written as

$$M_{++} = M_{--} = -\int d^3q \, \boldsymbol{F}(\mu_1)\boldsymbol{F}(\mu_2)\left[\left\{\xi_1 f_+^{(1)} + k^2 f_-^{(2)}\right\} + \xi_2 f_-^{(3)}\right], \tag{13}$$

$$M_{+-} = M_{-+} = -\int d^3q \, \boldsymbol{F}(\mu_1)\boldsymbol{F}(\mu_2)[\{f_-^{(1)} - k^2 f_-^{(2)}\}, \tag{14}$$

where the sign $+\ (-)$ in the M stands for right (left) circular polarized light, $\boldsymbol{F}$ represents the Fourier transform of the potential and $f_\pm^{(i)}$, $\xi_1$, and $\xi_2$ are functions of the internal and external momenta in terms of rational, logarithm and dilogarithm functions [10]. The magnitude square of the momentum of the two virtual photons is $\mu_1$ and $\mu_2$. The integral represents the virtual electron loop with momentum $\vec{q}$.

The Fourier transform, $\boldsymbol{F}(\mu)$, of the potential is similar to the electron case in Eq. (5). This function is finite and positive for small $\mu$ relative to the electron mass; at higher values the value of the function is small and highly oscillatory. This means that only small values of the internal momentum, $q \leq k$, in the integral in Eqs. (13) and (14) are significant. For wire diffraction at energies small compared to the electron mass, $k \ll mc^2$, we may use the approximations given in Ref. 10 for the functions in Eqs. (13,14); however, these approximations contain divergent terms of the momentum $\vec{q}$. Therefore, at any given order of $k$ one must make sure those singular terms in Eqs. (13,14) cancel.

## III. TWO BEAM CASE

In this section we explore wire diffraction when a wire is at the beam intersection as in Fig. 4. Two beams in phase cross each other at an angle $\alpha$.

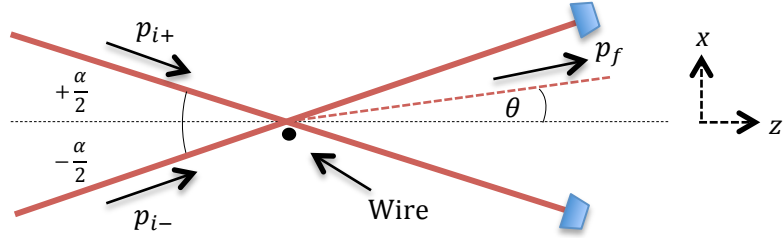

**FIG. 4. Beam interference and wire diffraction setup.** Two beams with a constant phase difference intersect at angle $\alpha$ and interfere. A wire is scanned across the beam intersection. Most particles go through unchanged. End detectors are set in front of each beam. The thin wire scatters particles. The scattering angle is $\theta$.

The 4-momentum vectors in Fig. 4 are:
$$p_{i\pm} = \left(\frac{E}{c}, \mp p \sin\frac{\alpha}{2}, 0, p \cos\frac{\alpha}{2}\right)$$
$$p_f = \left(\frac{E_f}{c}, p_f \sin\theta, p_{fy}, p_f \cos\theta\right).$$

Here we consider the electron case; the photon case is similar. The matrix element is the linear superposition:
$$S_{fi} = S_{fi+} + S_{fi-}, \tag{15}$$

where $S_{fi\pm} = -ie \int d^4x\, \bar{\psi}_f(\vec{x})\gamma^\mu A_\mu \psi_{i\pm}(\vec{x})$ and $\psi_{i\pm}(\vec{x}) = a u(p_{i\pm}, s_{i\pm}) e^{-ip_{i\pm}\cdot x}$. The superposition of amplitudes in Eq. (15) is interference in QED [12].

The magnitude of the momentum transfer is now $q_\pm^2 = 4p^2 \sin^2\left[\frac{\theta}{2} \pm \frac{\alpha}{4}\right]$. The transition probability per particle is calculated by standard methods used before in the paper; the result is
$$d\sigma = C\left|\bar{u}\gamma^0 u_- F_- + e^{i\Phi} \bar{u}\gamma^0 u_+ F_+\right|^2 d\theta, \tag{16}$$

where $F_\pm = {}_0F_1\left[2, -\frac{1}{4}q_\pm^2 R^2\right]$. We note that in order to apply arbitrary shifts to the interference pattern we have introduced the phase $\Phi$. At energies much smaller than the electron mass, the spin of the electron is unchanged during scattering and Eq. (16) becomes
$$\frac{d\sigma}{d\theta} = C\left(F_-^2 + 2F_- F_+ \cos\Phi + F_+^2\right), \tag{17}$$

where $C$ is a constant. In Fig. 5 we plot Eq. (17), evaluated at $\Phi=0$ (center of a bright fringe), together with Fraunhofer diffraction for a slit illuminated by the two intersecting beams. Once again Fraunhofer diffraction overestimates the wire radius.

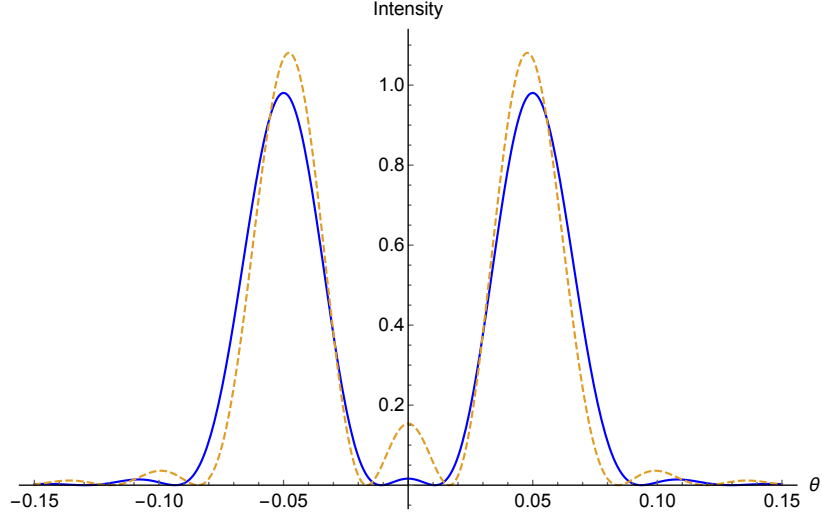

**FIG. 5. Diffraction pattern produced by wire illuminated by two beams that interfere constructively.** The quantum diffraction curve is in solid blue. Fraunhofer diffraction curve is in dashed orange. The beams intersect at $\alpha=0.1$. The wire is at the center of a bright fringe, which corresponds to $\Phi=0$. The wire diameter is 17 $\mu$m. The wavelength of the particle is 633 nm. Classical diffraction overestimates the wire radius.

We now observe the effect of the wire as we scan it across the beam intersection. We note that changing the phase $\Phi$ in Eq. (17) is equivalent to scanning the wire across the beam intersection while keeping the wire at the origin. In Fig. 6 we plot the probability distribution in Eq. (17) as a function of scattering angle $\theta$ and phase difference $\Phi$. As we scan the wire across the beam intersection we observe the presence of constructive and destructive interference. When the wire is at region of destructive interference the distribution reaches a minimum due to a decrease of diffracted photons.

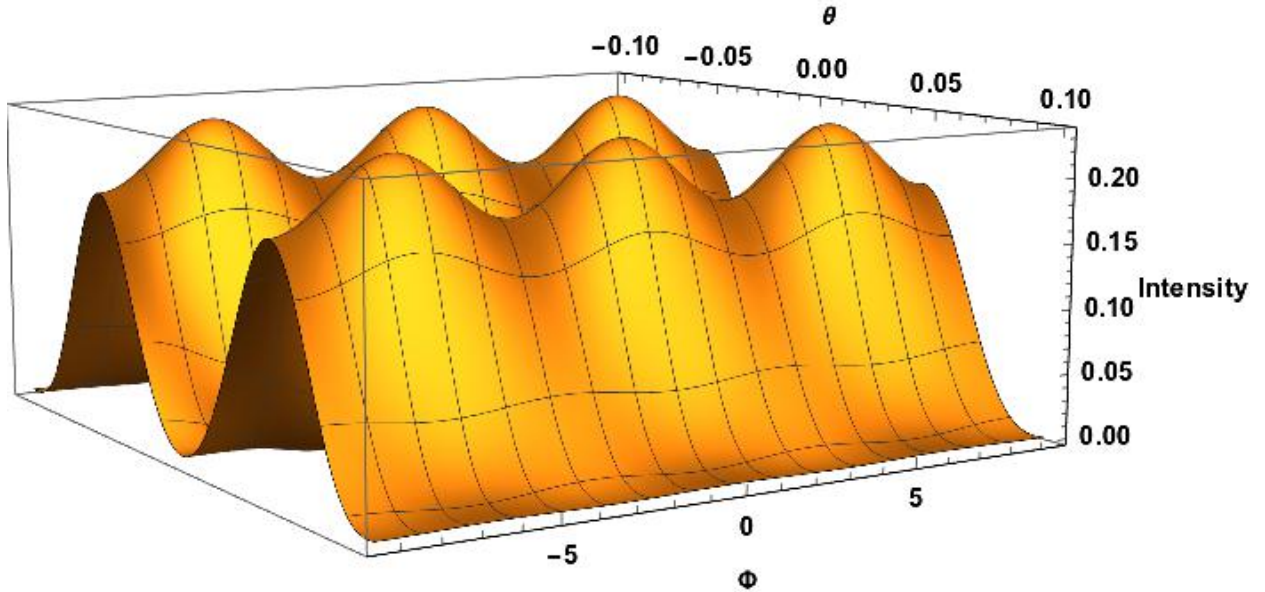

**FIG. 6. The wire is scanned across the beam intersection.** Two beams intersect at angle $\alpha=0.1$ and interfere. A wire, with 17 $\mu$m diameter, is at the beam intersection and produces interference. The wavelength of the particle is 633 nm. Scanning the wire across the beam intersection is equivalent to changing $\Phi$. We see a periodic increase and decrease of diffracted light as we reach a bright and dark fringe respectively. The angle $\theta$ is the scattering angle.

Wire diffraction with two interfering photon beams should result in a probability distribution similar to the electron case at the low energy level. An analysis of the incoming and outgoing photon momentum shows that the amplitude for the single beam of photons in Eqs. (13,14) may be used with one modification; the scattering angle $\theta$ changes

$$\frac{\theta}{2} \to \frac{\theta}{2} \pm \frac{\alpha}{4},$$

where the sign $\pm$ corresponds to the $\pm$ beam in Fig. 4. At the low energy limit, the probability distribution to maintain right circular polarization after scattering is

(18)
$$\frac{d\sigma_{++}}{d\theta} = C\left|(M_{++})_- + (M_{++})_+ e^{-i\Phi}\right|^2,$$

where $(M_{++})_\pm$ is the single beam amplitude in Eq. (13) that a right handed photon is scattered into a right handed photon evaluated at $\left(\frac{\theta}{2} \pm \frac{\alpha}{4}\right)$. An equation similar to Eq. (18) holds for $d\sigma_{+-}$.